\input{aipcheck}

\documentclass[
    ,final            
  ]
  {aipproc}

\layoutstyle{6x9}

\begin{document}

\title{Holographic Dark Energy: its Observational Constraints and Theoretical Features}


\author{Yin-Zhe Ma}{
  address={Institute of Astronomy,
University of Cambridge, Madingley Road, Cambridge, CB3 0HA, U.K.}}


\begin{abstract}
We investigate the observational signatures of the holographic dark
energy model in this paper, including both the original model and a
model with an interaction term between the dark energy and dark
matter. We first delineate the dynamical behavior of such models,
especially whether they would have a ``Big Rip'' for different
parameters, then we use several recent observational data to give
more reliable and tighter constraints on the models. The results
favor the equation of state of dark energy crossing $-1$, and the
universe ends in the "Big Rip" phase. By using the Bayesian evidence
as a model selection criterion to make the model comparison, we find
that the holographic
dark energy models are mildly favored by the observations compared with the $%
\mathrm{\Lambda CDM}$ model.
\end{abstract}

\maketitle


\section{Introduction}
The nature of dark matter (DM) and dark energy (DE), which
constitute about 95\% of the total cosmic density, are among the
most important problems in modern physics and astronomy
\cite{coincidence}. Dark energy may be a problem that has to be
solved in the framework of fundamental microscopic physics such as
string theory \cite{Witten:2000zk}.

\begin{figure}[t]
\centerline{\includegraphics[bb=0 0 356 235,width=2.8in]{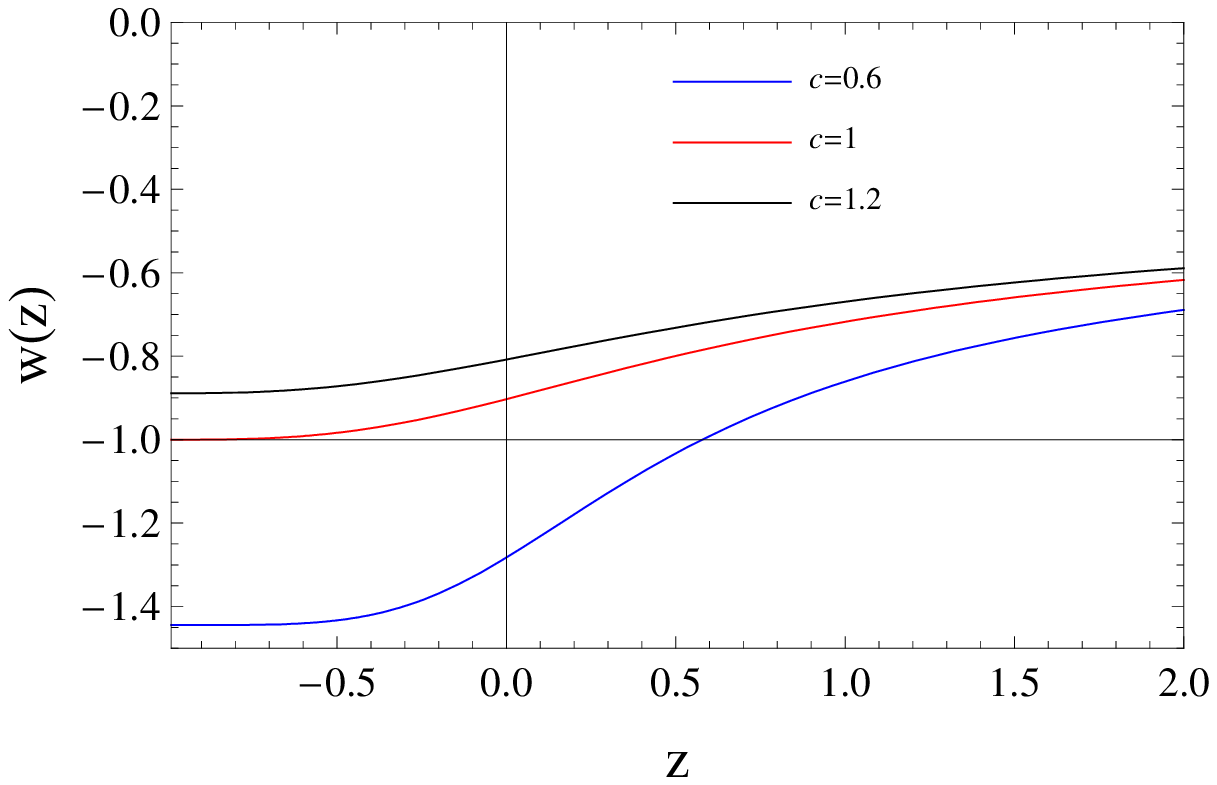}
\includegraphics[bb=0 0 401 266,width=2.7in]{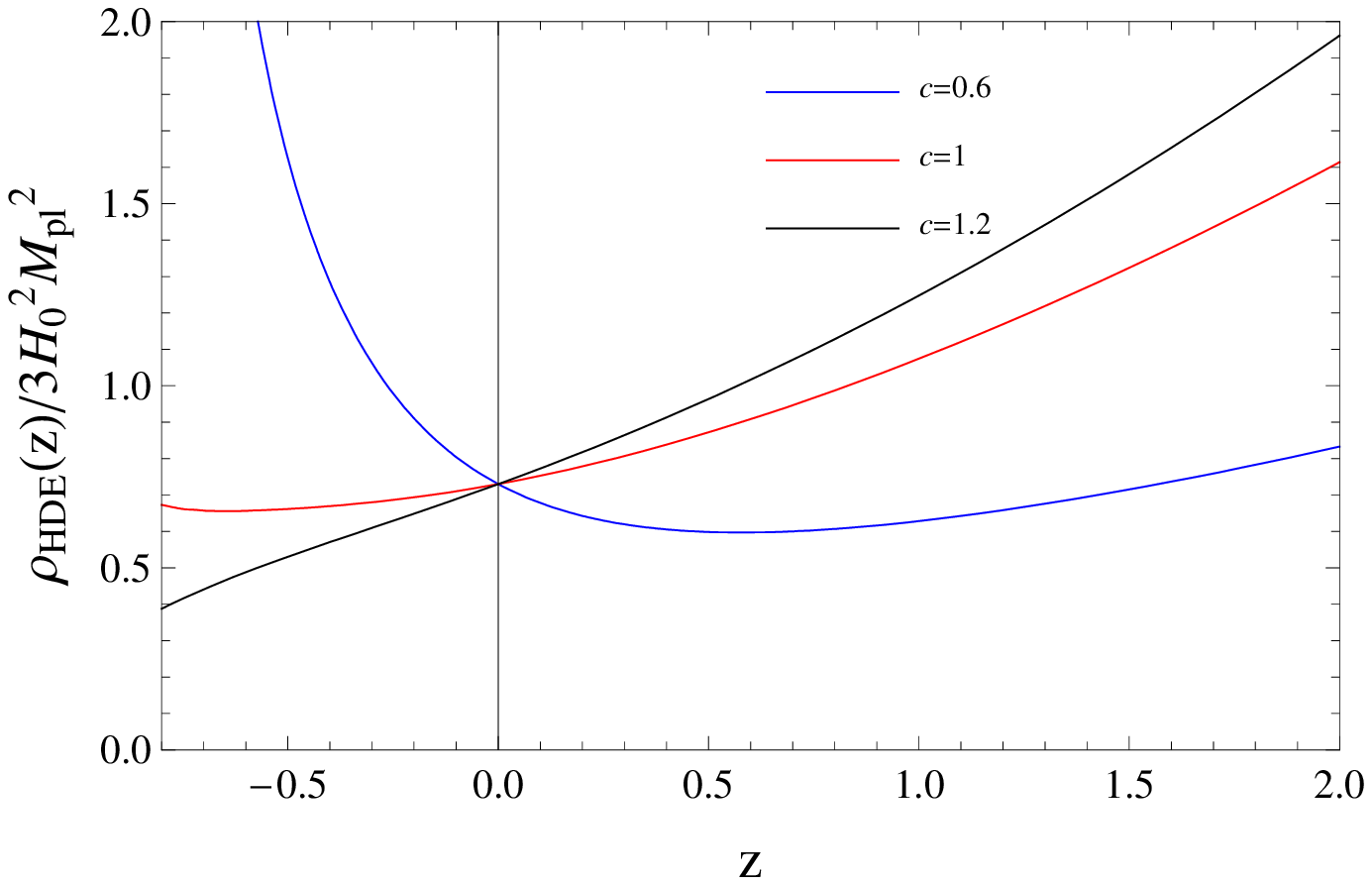}}
\caption{EEoS and HDE density in the holographic dark energy model
for different value of $c$. Left: EEoS; Right: HDE density. Here we
set $\Omega _{m0}=0.27$.} \label{EEoS1}
\end{figure}

Here we consider the dark matter-dark energy interaction in the case
of the so-called holographic dark energy (HDE) model
\cite{Cohen:1998zx,Horava:2000tb,Hsu:2004ri,Li:2004rb,ma07}. This is
based on the holographic principle \cite{holoprin}, which was
inspired by the Bekenstein entropy bound of black holes \cite{bh}.
If the quantum zero-point energy density $\rho _{\Lambda }$ is
relevant to an ultraviolet cut-off, the total energy of the whole
system with size $L$
should not exceed the mass of a black hole of the same size, thus we have $%
L^{3}\rho _{\Lambda }\leq LM_{\mathrm{pl}}^{2}$. The largest
infrared cut-off $L$ is chosen by saturating the inequality so that
the HDE density is
\begin{equation}
\rho _{\mathrm{de}}=3c^{2}M_{\mathrm{pl}}^{2}L^{-2}~,
\label{holo-density}
\end{equation}%
where $c$ is a numerical constant, and $M_{\mathrm{pl}}\equiv
1/\sqrt{8\pi G} $ is the reduced Planck mass. $L$ should be the size
of the future event horizon, since only this case would result in a
dark energy component which drives the accelerated expansion of the
universe \cite{Li:2004rb}, i.e.
\begin{equation}
L=R_{eh}(t)=a(t)\int_{a}^{\infty }\frac{da^{^{\prime }}}{H^{^{\prime
}}a^{^{\prime }2}}.
\end{equation}
If we consider the dark matter and dark energy interaction, the
energy conservation equation will be written as
\begin{equation}
\dot{\rho}_{m}+3H\rho_{m}=Q,  \label{Matter}
\end{equation}%
\begin{equation}
\dot{\rho}_{\mathrm{de}}+3H(1+w_{\mathrm{de}})\rho
_{\mathrm{de}}=-Q, \label{Dark Energy}
\end{equation}%
where $\rho_{m}$, $\rho_{\mathrm{de}}$ and $w_{\mathrm{de}}$ are
matter density, HDE density and HDE equation of state respectively.
Here we consider the physically plausible interaction as $Q=3\alpha
H\rho _{\mathrm{de}}$ ($\alpha$ is a parameter describing the
strength of interaction), which has been widely used for the
interaction between the matter and massive scalar field
\cite{Ferreira}. Therefore, by combining the above equations and
Friedmann equation, we quickly obtain the following equations which
describe the evolution of the fractional energy density
$\Omega_{\mathrm{de}}$, Hubble parameter $H(z)$, and also the HDE
effective equation of state (EEoS) (see \cite{ma07} for details):
\begin{equation}
\frac{d\Omega _{\mathrm{de}}(z)}{dz}+\frac{\Omega _{\mathrm{de}}}{1+z}%
[(1-\Omega _{\mathrm{de}})(1+\frac{2}{c}\sqrt{\Omega _{\mathrm{de}}}%
)-3\alpha \Omega _{\mathrm{de}}]=0,  \label{Density 2}
\end{equation}%
\begin{equation}
\frac{dH}{dz}=-\frac{H(z)}{1+z}[\frac{1}{2}\Omega _{\mathrm{de}}(1+3\alpha +%
\frac{2}{c}\sqrt{\Omega _{\mathrm{de}}})-\frac{3}{2}], \label{Hubble
3}
\end{equation}%
\begin{equation}
w_{\mathrm{eff}}(z)=w_{de}+\alpha =-\frac{1}{3}-\frac{2}{3}\frac{\sqrt{%
\Omega _{\mathrm{de}}}}{c}.  \label{EEoS}
\end{equation}%
By solving the Eqs. (\ref{Density 2}), (\ref{Hubble 3}), and
(\ref{EEoS}), we can know the dynamics of the HDE model as follows:
For the non-interacting HDE model, i.e. $\alpha=0$, the evolution of
HDE density and its EEoS will be affected by the value of parameter
$c$ as shown in Fig. \ref{EEoS1}.
If $c=1$, the EEoS of HDE will turn out to be $-1$ (Cosmological
Constant), and the universe will end up in a de-Sitter phase. If
$c>1$, the EEoS is always greater than $-1$, and the HDE is just
like the quintessence dark energy. On the contrary, if $c<1$, the
EEoS will cross $-1$ at some time, and the universe will end up in a
"Big Rip", indicating the HDE resembles the behavior of phantom dark
energy \cite{Caldwell99}.

For the interacting HDE model (IHDE), we vary the value of $\alpha$
but fix $c=1$ to see the interacting effect.
\begin{figure}[t]
\centerline{\includegraphics[bb=0 0 384 252,width=2.8in]{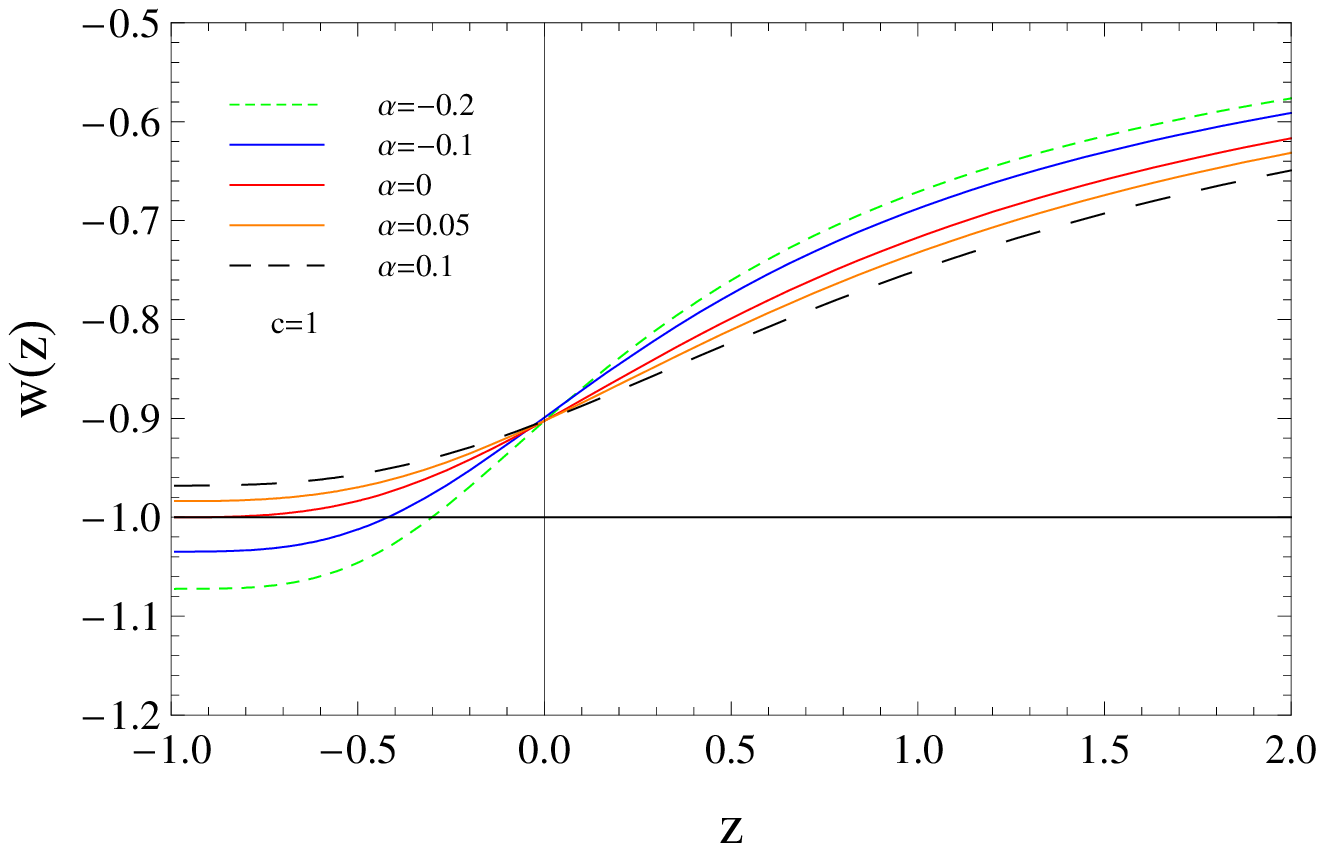}
\includegraphics[bb=0 0 355 241,width=2.7in]{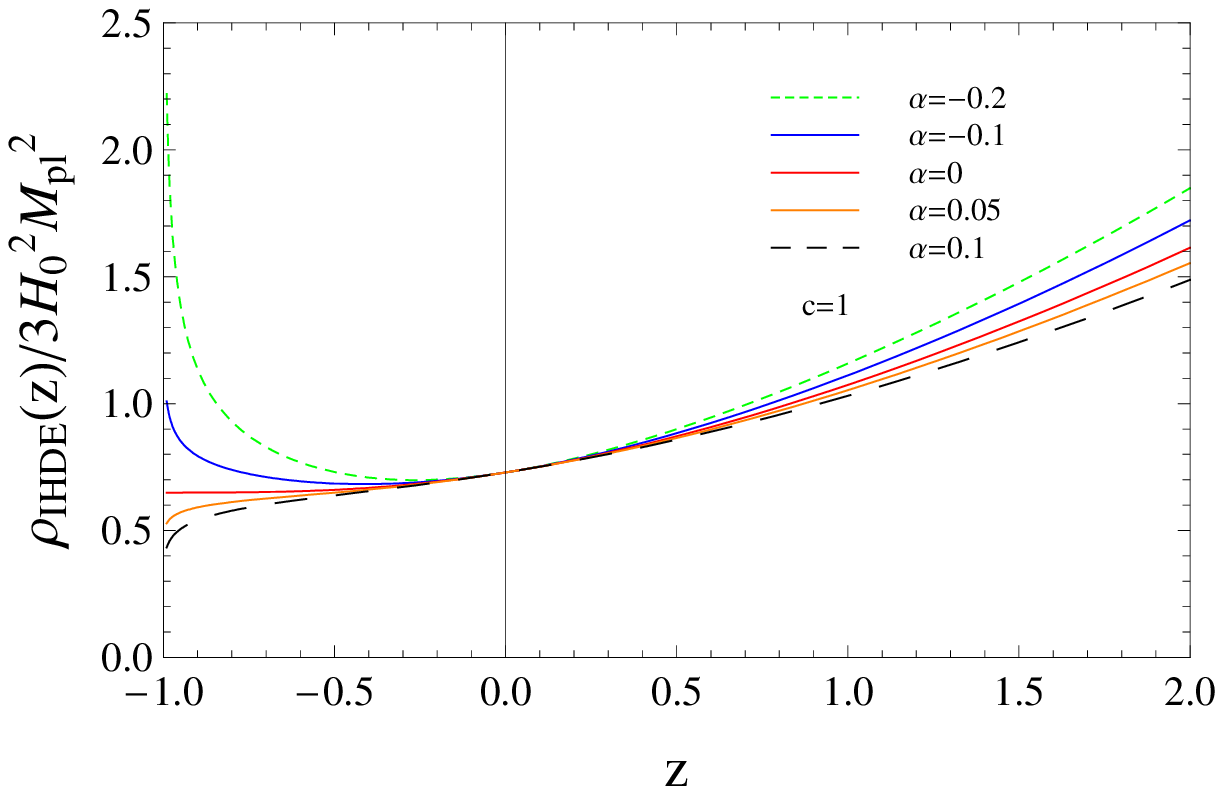}}
\caption{EEoS and IHDE density in the interacting holographic dark
energy model for different value of $\alpha$. Left: EEoS; Right: HDE
density. Here we set $\Omega _{m0}=0.27$.} \label{EEoS2}
\end{figure}
In Fig. \ref{EEoS2}, we see that if $\alpha<0$, the dark matter
"decays" into IHDE (Eqs. (\ref{Matter}) and (\ref{Dark Energy})),
which will make the HDE density increase rapidly at the later era of
cosmic evolution. Thus, effectively the IHDE will resemble the
phantom dark energy with EEoS less than $-1$ in the near future (see
the left of Fig. \ref{EEoS2}). On the other hand, if IHDE "decays"
into dark matter, i.e. $\alpha>0$, the IHDE density would decay more
rapidly, making the IHDE behave like a quintessence field.

However, we want to ask the inverse questions: what values of
parameter $c$ and $\alpha$ are preferred by the current
observational data? In addition, which of these dynamic dark energy
models are favored by current observational data compared with the
concordance $\Lambda$CDM model?
\section{Methodology}
We utilize several data sets to constrain the parameters of the HDE
and IHDE model, including the 182 high-quality type Ia supernovae
\cite{Riess06}, the baryon acoustic oscillation measurement from the
Sloan Digital Sky Survey \cite{Eisenstein05}, 42 latest X-ray gas
mass fraction data from Chandra observations \cite{Allen07}, 27 GRB
samples generated with $E_{peak}-E_{\gamma }$ correlation
\cite{Schaefer07}, and the CMB shift parameter from the WMAP 3 year
results \cite{wang07}. To break the degeneracy and explore the power
and differences of the constraints for these data sets, we use them
in several combinations to
perform our fitting: $\mathrm{SN_{sel}+BAO,SN_{sel}+BAO+f_{gas}}$, and $%
\mathrm{SN_{sel}+BAO+f_{gas}+GRB+CMB}$.

For comparing different models, one must choose a statistical variable. The $%
\chi _{min}^{2}$ is the simplest one and is widely used. However,
for models with different numbers of parameters, the comparison
using $\chi^2$ may not be fair, as one would expect that models with
more parameters tend to have lower $\chi^2$. Instead, we use the
Bayesian evidence (BE) as a model selection criterion. The Bayesian
evidence of a model $M$ takes the form
\begin{equation}
\mathrm{BE}=\int {\mathcal{L}(\mathbf{d|\theta
},M)\mathbf{p}(\mathbf{\theta }|M)d\mathbf{\theta }},  \label{BE}
\end{equation}%
where $\mathcal{L}(\mathbf{d|\theta },M)$ is the likelihood function
given
the model $M$ and parameters $\mathbf{\theta }$, and $\mathbf{p}(\mathbf{%
\theta }|M)$ is the priors of the parameters. The BE may be the best
model selection criterion, as it is the average of likelihood of a
model over its prior of the parameter space and automatically
includes the penalties of the number of parameters and data, so it
is more direct, reasonable and
unambiguous than the $\chi _{min}^{2}$ model selection \cite%
{Liddle06}. The logarithm of BE can be used as a guide for model
comparison, and we choose the $\Lambda $CDM as the reference
model: $\Delta \ln \mathrm{BE}=\ln \mathrm{BE}_{model}-\ln \mathrm{BE}%
_{\Lambda CDM} $. The strength of the evidence for the model is
considered according to the numerical value of BE: $\Delta \ln
\mathrm{BE}<1$ (Weak); $1< \Delta \ln \mathrm{BE}<2.5$
(Significant); $2.5<\Delta \ln \mathrm{BE}<5$ (Strong to very
strong); $\Delta \ln \mathrm{BE}>5 $ (Decisive).
We use the nested sampling technique to compute BE \cite%
{Mukherjee06,Skilling}.
\section{Results}
In Table \ref{table2}, we give the best fits and the $1\sigma $ CL
of the HDE model parameters, as well as the value of $\ln \Delta
\mathrm{BE}$ for the three data set combinations. The best fit of
$c$ varies slightly across the different data sets, it is 0.761 for
the SN+BAO data set, but decreases slightly when the
$f_{\mathrm{gas}}$, GRB and CMB data are included. However, for all
data sets, we have $c<1$ at more than $1.5\sigma $, indicating that
the EEoS of HDE will cross $-1$ at some redshift, and the energy
density of HDE will diverge in the future (see the blue curve in
Fig. \ref{EEoS1}).
\begin{table}[hb]
\caption{The fitting result for the HDE model.}%
\begin{tabular}[t]{c|c|c|c}
 & $\rm SN+BAO$ & $\rm SN+BAO+f_{gas}$ & $\rm SN+BAO+f_{gas}+GRB+CMB$\\
\hline
$\Omega_{m0}$ & $0.273^{+0.020}_{-0.020}$ & $0.270^{+0.021}_{-0.018}$ & $0.276^{+0.017}_{-0.016}$ \\
$c$ & $0.761^{+0.154}_{-0.117}$ & $0.745^{+0.130}_{-0.101}$ &
$0.748^{+0.108}_{-0.093}$ \\
$\Delta\ln{\rm BE}$ &$~0.09\pm0.12~$ & $~0.63\pm 0.18~$&$~0.65\pm0.18~$\\
\end{tabular}
\label{table2}
\end{table}
The HDE model fits about equally well ($\ln$ BE=0.09) as the
$\Lambda$CDM when we only use the SNIa and BAO data. With $f_{gas}$,
GRB and CMB data added, it fits mildly better than the $\Lambda$CDM,
but with the data presently available the difference is not
significant ($\ln$ BE =$0.63\sim 0.65$).

In Table \ref{table3}, we show the fitting results for the IHDE
model.
\begin{table}[h]
\caption{The Fitting results the IHDE model.}%
\begin{tabular}[t]{c|c|c|c}
 & $\rm SN+BAO$ & $\rm SN+BAO+f_{gas}$ & $\rm SN+BAO+f_{gas}+GRB+CMB$\\
\hline
 $\Omega_{m0}$ & $0.272^{+0.023}_{-0.022}$ & $0.275^{+0.021}_{-0.021}$ & $0.281^{+0.017}_{-0.017}$ \\
c & $0.592^{+0.204}_{-0.113}$ & $0.667^{+0.321}_{-0.164}$ & $0.692^{+0.135}_{-0.107}$\\
 $\alpha$ & $-0.020^{+0.145}_{-0.174}$ & $0.068^{+0.093}_{-0.120}$ & $-0.006^{+0.021}_{-0.024}$ \\
$\Delta \ln \mathrm{BE}$& $~0.41 \pm 0.12~$ &$ ~0.70\pm0.18~$& $ ~0.75\pm0.18~$\\
\end{tabular}
\label{table3}
\end{table}
\begin{figure}[b]
\centerline{\includegraphics[bb=0 0 410
263,width=2.8in]{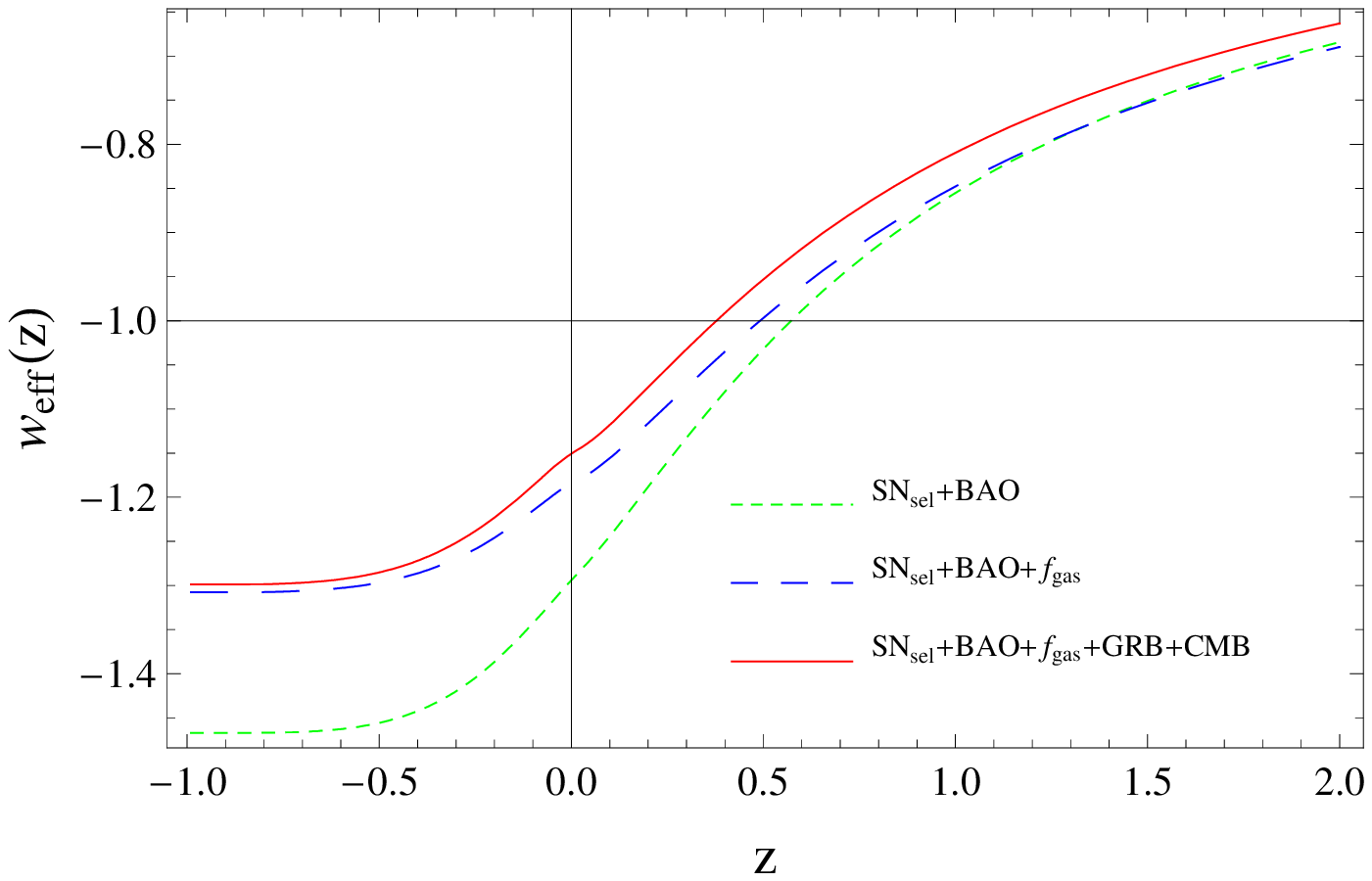}
\includegraphics[bb=0 0 370 241,width=2.7in]{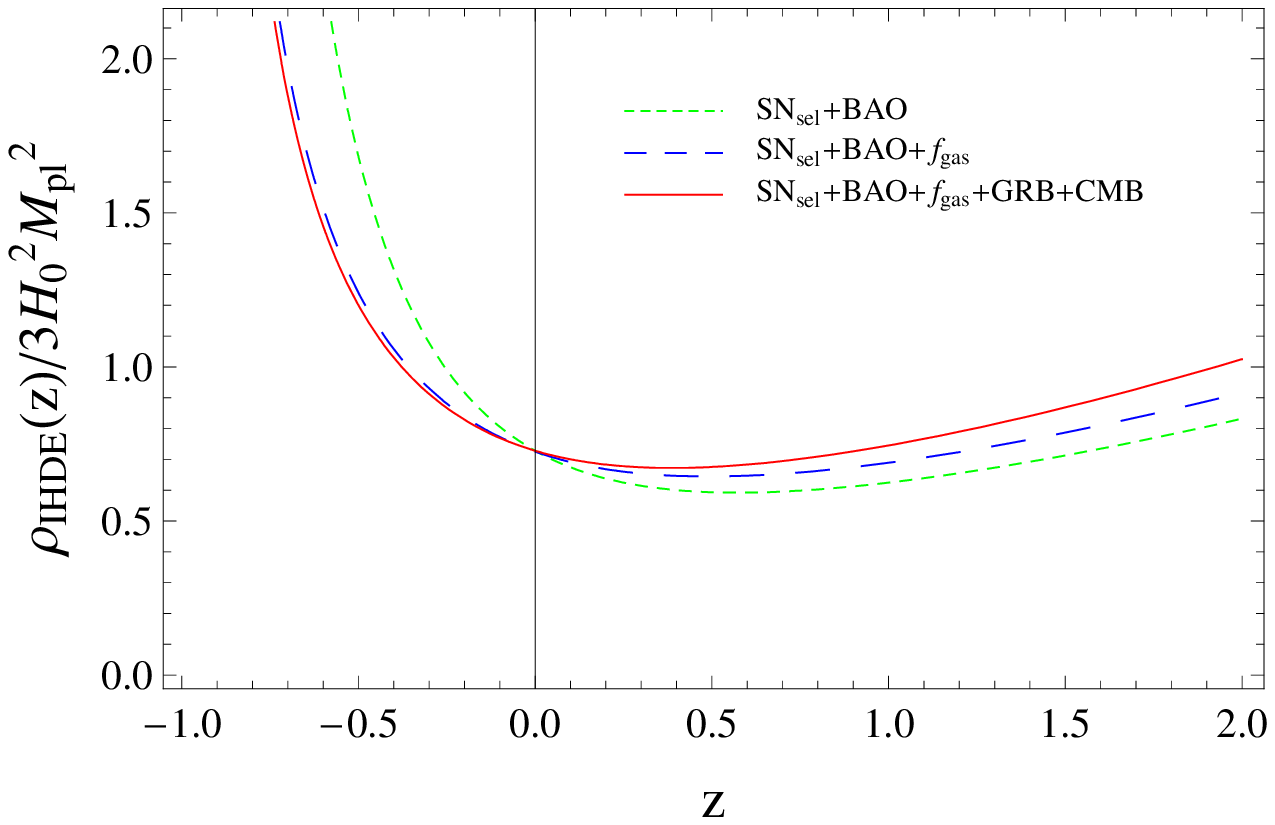}
} \caption{EEoS and energy density evolution for the best fit IHDE
models.} \label{fig:IHDE}
\end{figure}
From the best fits and $1\sigma$ CL, the fractional energy density
of dark matter is still around 0.27, and the parameter $c$ is always
less than $1$. The difference from the HDE model is that IHDE has
the interacting parameter $\alpha$ here and its value is either
positive or negative, but always very small and around zero. If you
look closer at it, zero is always covered in the $1\sigma$ CL,
indicating that there is little evidence for the interaction.
Furthermore, the Bayesian Evidence is always positive but less than
$1$, if you compare them with the previous Bayesian Evidence of
non-interacting case, you can see that the number is slightly
greater than the previous case, which means that the IHDE model is a
little more favored than the HDE model but the evidence is still not
strong.

Using the best fitting values for the three data sets, we also plot
the effective equation of state and dark energy density in Fig.
\ref{fig:IHDE}. We can see that all three data sets suggest the
equation of state crosses $-1$ and the universe ends in the "Big
Rip" in the future, just like the HDE model.

\begin{figure}
  \includegraphics[bb=7 157 571 701,height=2.4in]{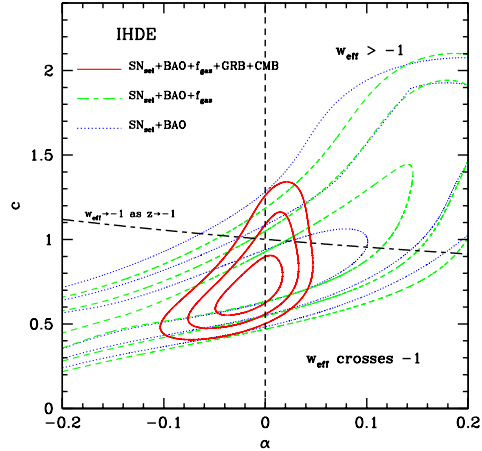}
  \caption{The contours of c vs. $\alpha$ in the IHDE model.}
\label{contouralphac}
\end{figure}

The contour map for $c$ and $\alpha$ is plotted in Fig.
\ref{contouralphac}. In Fig. \ref{contouralphac}, the contours
corresponding to the data sets SN+BAO+${f_{gas}}$+GRB+CMB are much
tighter than the other data sets. This is because we use the data
set GRB and CMB, which has very large redshift distribution so they
break the degeneracy of the parameter $c$ and $\alpha$. We also mark
the $\alpha-c$ values for which $w_{eff}=-1$ as a dashed line. This
forms the dividing line between quintessence-like and phantom-like
behavior. For the best fit parameters of all three data set
combinations, $c \sim 0.6 $, so $c<1$. The value of $\alpha$ varies
more, but all consistent with being 0 within $1.5\sigma$, which
suggests the evidence for the interaction is very weak. In any case,
for all three data set combinations, the best fit value resides in
the phantom-like region, although a large area of quintessence-like
region is also allowed.


\section{Conclusion}
In this paper we first introduced the holographic dark energy model
with the interacting term $Q=3\alpha H\rho _{\mathrm{de}}$, so the
non-interacting case could be viewed as the special case with
$\alpha =0$. We illustrated the dynamical behavior of these models
by choosing some representative values of the parameters $c$ and
$\alpha$. The condition for the model to have a ``Big Rip'' is
determined. Second, we utilize several data sets from the recent
observations to constrain the models. The best-fits for the three
data sets are given in Table \ref{table2} for the HDE model, and
Table \ref{table3} for the IHDE model. For both the HDE and IHDE
models, the data favors ``phantom'' behavior slightly, i.e. the dark
energy initially has $w_{eff}>-1$, but eventually crossing the
phantom dividing line, and the model ends with a ``Big Rip''.
However, quintessence-like behavior is also allowed with the present
data. Next, we utilize the Bayesian evidence (BE) as a model
selection criterion to compare the holographic models with the
$\Lambda $CDM model for the three data sets. Both the HDE and the
IHDE model are mildly favored by the current observational data set,
although the evidence is weak.

In brief, we conclude that according to the observational data, the
holographic dark energy model, especially the interacting
holographic dark energy model is mildly favored by the current data,
and for the best fit model the equation of state for both the HDE
and IHDE crosses $-1$, for which the Universe ends up in a "Big
Rip".


\begin{theacknowledgments}
I acknowledge the support from the Cambridge Overseas Trust and a
studentship from Trinity College, Cambridge, as well as the National
Science Foundation in China. I also thank Xuelei Chen, Yan Gong and
many other colleagues for the helpful discussions, and David Cline
for inviting me to give this talk.
\end{theacknowledgments}



\bibliographystyle{aipproc}   


\IfFileExists{\jobname.bbl}{}
 {\typeout{}
  \typeout{******************************************}
  \typeout{** Please run "bibtex \jobname" to optain}
  \typeout{** the bibliography and then re-run LaTeX}
  \typeout{** twice to fix the references!}
  \typeout{******************************************}
  \typeout{}
 }


\end{document}